\documentclass[12pt]{article}
\begin{document}
\title{\bf Spherically symmetric solution of the Weyl--Dirac theory of gravitation and possible influence of dark matter on the interplanetary spacecraft motion}   
\author{O. V. Babourova, B. N. Frolov,  P. E. Kudlaev, E. V. Romanova \\
Moscow Pedagogical State  University,\\
Institute of Physics, Technology and Information Systems,\\
M. Pirogovskaya ul. 29, Moscow 119992, Russian Federation}

\date{}
\maketitle
\renewcommand{\abstractname}{Abstract}
\begin{abstract}
The Poincar{'}e and Poincar{'}e--Weyl gauge theories of gravitation  with Lagrangians quadratic on curvature and torsion in post-Riemannian spaces with the Dirac scalar field is discussed in a historical aspect. The various  hypothesizes concerning the models of a dark matter with the help of a scalar field are considered. The new conformal Weyl--Dirac theory of gravitation is proposed, which is a gravitational theory in Cartan--Weyl spacetime with the Dirac scalar field  representing the dark matter model. A static spherically symmetric solution of the field equations in vacuum for a central compact mass is obtained as the metrics conformal to the Yilmaz--Rosen metrics. On the base of this solution one considers a radial movement of an interplanetary spacecraft starting from the Earth. Using the Newton approximation one  obtains that the asymptotic line-of-sight velocity in this case depends from the parameters of the solution, and therefore one can obtain on basis of the observable data the values of these parameters.  
\end{abstract}
{\bf Keywords:} Dark matter, Dirac scalar field, Weyl--Dirac theory of gravitation, Cartan--Weyl spacetime, Yilmaz--Rosen metrics, spacecraft Earth flyby

\section{Introduction}

In \cite {Kib} a gauge principle has  been applied to Poincar{'}e  group that has resulted in construction a gauge theory of gravitation in a post-Riemannian space with curvature and torsion -- a Riemann--Cartan space.  In \cite {Kib} a curvature scalar (generalized on a Riemann--Cartan space) were used as a Lagrangian of the theory. The such generalized theory of gravitation has named as  the Einstein--Cartan theory of gravitation. The similar gauge theory of gravitation was advanced in \cite {Fr1, Fr3, Fr4, Fr5} where it was offered to use as a Lagrangian along with a curvature  scalar also quantities quadratic on  curvature and torsion.  Later in \cite {Hay} the most general Lagrangian of a such kind in a Riemann-Cartan space was constructed containing  ten arbitrary connection constants. A gauge theory of gravitation with quadratic  Lagrangians has received the name  the Poincar{'}e  gauge theory of gravitation, see  Refs. in \cite {Hehl_rew1, Fr_book, Hehl_book}.
 
Then in \cite {Fr_conf2, Fr_GPS} it was advanced a conformally invariant generalization  of the Poincar{'}e  gauge theory of gravitation with use of the method offered  by Dirac in his well-known work \cite {Dirac}. The given method is based on using in a Lagrangian of the theory an additional scalar field which in \cite {Fr_conf2} was named as a Dirac scalar field. 

In  \cite {BFZ1, BFZ3} the gauge theory of gravitation was constructed, proceeding from the requirement of the gauge invariance of the theory concerning the Poincar{'}e--Weyl  group, supplementing  the Poincar{'}e group by a group of spacetime stretching and compression (dilatations). It was shown that from this requirement  spacetime obtains a geometrical structure of Cartan--Weyl  space. Besides,  in this theory a requirement  appears of necessary existence of the additional scalar field having so fundamental geometrical status, as well as the metrics. The further development of the theory has shown, that the given scalar field coincides by the properties with the scalar field entered by Dirac in  \cite {Dirac}. 

Further on the basis of the given result the theory of gravitation in a Cartan--Weyl  space  was constructed \cite {BabFr_ArX1, BFK_ArX2, BFKost, BabFr_mon}, which generalizes the Einstein--Cartan and the Poincar{'}e  gauge theories of gravitation on presence of nonmetricity of the Weyl's type and uses the Dirac scalar field for supporting  conformance of the theory. This generalized theory of gravitation is pertinent to name {\em the Weyl--Dirac theory of gravitation}.
 
According to E. B. Gliner \cite {Gliner}, the cosmological constant in the Einstein equation  determines  a vacuum energy density (dark energy). In the Weyl--Dirac  conformal  theory of gravitation, an effective cosmological constant appears, which value is determined  by the Dirac scalar field. Application of the Weyl--Dirac theory to the early universe cosmology has allowed to find the solution of a well-known cosmological constant problem \cite {BabFr_mon, BFL_IzvVuz, BFL_GrCos, BF_ResInt}, which represents the important problem of modern physics \cite {Wein, China}. 

Matos with co-authors \cite {Matos_IntJ}  within the framework of a Riemann geometry in GRG was advanced a cosmological SFDM model, in which the dark matter is modelled with the help of a scalar field using a special kind of a potential/ The full solution of a cosmological scenario was obtained. To the hypothesis that the scalar field can carry out the same problems, which are assigned to a dark matter, has joined Capozziello with coauthors \cite {Cappozz1}. In \cite {Cappozz1} it is used Yukawa  interaction between the scalar field and substance.  

In the monograph \cite {BabFr_mon} within the framework of the Weyl--Dirac theory of gravitation, the hypothesis was stated that the Dirac scalar field in Cartan--Weyl  space not only determines a size of the effective cosmological constant (dark energy density), but also plays a role the basic components of a dark matter. Then the spherically symmetric solution of the Weyl--Dirac theory for the central mass in vacuum \cite {BabFr_mon, BFFeb_IzvVuz1, BFKudRom} was found. 
 
Thus, the hypothesis about a possible modeling of a dark matter by a scalar field is fruitful idea, which now is developed by some modern researchers. In the given work the new method of deriving the spherically symmetric solution of the conformal Weyl--Dirac theory of gravitation will be elaborated, and also possible influence of a dark matter on movement of space vehicles within the limits of Solar system is found out. 

\section{Lagrangian density and field equations in Weyl--Dirac theory} 
 
Let us consider \cite{BabFr_mon} a connected 4D oriented differentiable manifold $\mathcal M$ equipped with a metric $\hat{g}$ of the index 3, a linear connection $\Gamma^{a}{}_{b}$ and a volume 4-form $\eta $. Then a Cartan--Weyl space $\mathcal CW_4$ is defined as the such manifold equipped with a curvature 2-form ${\mathcal R}^{a}{}_{b}$, a  torsion 2-form ${\mathcal T}^{a}$ and a nonmetricity 1-form ${\mathcal Q}_{ab}$ obeing the Weyl condition
\begin{equation}
\mathcal Q_{ab} =\frac{1}{4} g_{ab} \mathcal Q\,. \label{eq:weylcond} 
\end{equation}
Here ${\mathcal Q}_{ab}=-\mathcal D g_{ab}$, and $\mathcal D = {\rm d} + \Gamma \wedge \dots$ is the exterior covariant differential. 

In \cite{BFZ1, BFZ3} the Poincar\'{e}--Weyl gauge theory of gravitation (PWTG) has been developed. The gauge field introduced by the subgroup of dilatations is   named as dilatation field, its vector-potential is the Weyl 1-form, and quanta of this field can have nonzero rest mass. An additional scalar field $\beta(x)$ is introduced in PWTG as an essential geometrical addendum to the metric tensor, the tangent space metrics being the form,
\begin{equation}
\label{eq:tangmetr}
g_{ab} = \beta^{-2} g^M_{ab}\,, 
\end{equation}
where $g^M_{ab}$ are the constant components of the Minkowski metric tensor. 

The properties of the field $\beta(x)$ coincide with those of the scalar field introduced by Dirac \cite{Dirac}. Some terms of the Dirac scalar field Lagrangian have structure of the Higgs Lagrangian and can cause an appearance of nonzero rest masses of particles \cite{Fr_GPS}. 

On the basis of PWTG, the conformal theory of gravitation in Cartan--Weyl spacetime with Dirac scalar has been developed \cite{BabFr_ArX1, BFK_ArX2, BFKost, BabFr_mon} with the Lagrangian density 4-form (in exterior form formalism) \cite{BFL_IzvVuz, BFL_GrCos},
\begin{eqnarray}
\mathcal L &=& \mathcal L_G + {\mathcal L}_{mat} + \beta ^{4} \Lambda ^{ab} \wedge \left (\mathcal Q_{ab} -\frac{1}{4} g_{ab} \mathcal Q \right )\,, \label{eq:LagrTot}\\
\mathcal L_G &=&2f_{0} \left[ \frac{1}{2} \beta ^{2} \mathcal R^{a}{} _{b} \wedge \eta _{a}{} ^{b} -\beta ^{4} \Lambda \eta +\frac{1}{4} \lambda \mathcal R^{a}{} _{a} \wedge \ast\mathcal R^{b}{} _{b} +\right. \nonumber\\ 
&& + \rho _{1} \beta ^{2} \mathcal T^{a} \wedge \ast\mathcal T_{a} +\rho _{2} \beta ^{2} (\mathcal T^{a} \wedge \theta _{b} )\wedge \ast(\mathcal T^{b} \wedge \theta _{a} )+ \nonumber\\ 
&& +\rho _{3} \beta ^{2} (\mathcal T^{a} \wedge \theta _{a} )\wedge \ast(\mathcal T^{b} \wedge \theta _{b} )+\xi \beta ^{2} \mathcal Q\wedge \ast\mathcal Q+\zeta \beta ^{2} \mathcal Q\wedge \theta ^{a} \wedge \ast\mathcal T_{a} + \nonumber\\
&& + l_{1} \rm d\beta\wedge\ast \rm d \beta + {\it l}_{2} \beta \rm d \beta \wedge \theta^{\it a} \wedge \ast\mathcal T_{\it a} \left. + {\it l}_{3} \beta \rm d \beta\wedge \ast\mathcal Q \frac{}{} \right]\, .\label{eq:LagrPL2}
\end{eqnarray}
Here $\mathcal L_G$ is the gravitational field Lagrange density, ${\mathcal L}_{mat}$ is the matter Lagrange density. The first term in $\mathcal L_G$ is the Gilbert--Einstein Lagrangian density generalized to the Cartan--Weyl space ($\eta_a{}^b=\ast(\theta_a\wedge\theta^b$), $f_0 = c^4/16\pi G$), the second term is a generalized cosmological term describing a vacuum energy ($\Lambda$ is the Einstein cosmological constant).  

We use the exterior form variational formalism on the base of the Lemma on the commutation rule between variation and Hodge star dualization \cite{BFK_CQG}. The independent variables are the nonholonomic connection 1-form $\Gamma^{a}{}_{b} $, the basis 1-form $\theta ^{a} $, the Dirac scalar field $\beta(x)$, and the Lagrange multipliers $\Lambda ^{ab}$. $\Lambda$--equation  yields the Weyl's condition (\ref{eq:weylcond}). The variational field equations of the theory ($\Gamma$-equation, $\theta$-equation and  $\beta$-equation) can be found in \cite{BFL_GrCos, BabFr_mon}. 

These variational field equations have been solved at the very early stage of evolution of universe for the scale factor $a(t)$ and the field $\beta (t)$, when a matter density was very small \cite{BabFr_mon, BFL_IzvVuz, BFL_GrCos, BF_ResInt}. This solution realizes  exponential diminution of the field $ \beta $, and thus sharp exponential decrease of physical vacuum energy (dark energy) by many orders.  Thus this result can explain the exponential decrease in time at very early Universe of the dark energy  describing by the effective cosmological constant. This can give a way to solving the one of the fundamental problems of the modern theoretical physics -- the problem of cosmological constant  (see \cite{Wein, China}) -- as a consequence of fields dynamics at the early Universe. 

\section{Spherically symmetric solution \\ of the Weyl--Dirac theory}

Now a static spherically symmetric solution of the field equations in vacuum (in case of $\Lambda =0, \; \lambda =0$) is obtained  for a central compact mass $m$ \cite{BabFr_mon, BFFeb_IzvVuz1, BFKudRom}. 

In the spherically symmetric case the torsion 2-form is, 
${\mathcal T}^{a}=(1/3)\mathcal T \wedge \theta^{a}\qquad \mathcal T = *(\theta _{a} \wedge *\mathcal T^{a})$ , were $\mathcal T $ is a torsion trace 1-form.

As a consequence of the $\Gamma$-equation, one can conclude that the torsion 1-form $\mathcal T$ and the nonmetricity 1-form $\mathcal Q $ can be realized as $\mathcal T = s{\rm d}U \, , \qquad \mathcal Q = q{\rm d}U\,, \qquad   U=\log \beta$ , where  $s$ and $q$ are arbitrary numbers.

We shall find a static spherically symmetric solution with a metrics of the form,  
\begin{equation}
\label{eq:metr}
ds^2= e^{\lambda(r)-\mu(r)}dt^2-e^{\lambda(r)+\mu(r)}
(dr^2+r^2(d\theta^2+\sin^2{\theta}\,d\phi^2))\,. 
\end{equation}
After calculation the $\Gamma$-, $\theta$- and $\beta$-equations with the help of this metrics, one can conclude that these equations are reduced to the following equations, 
\begin{eqnarray}
&&\mu''+\frac{2}{r}\mu'=0\,,\qquad \lambda=- 2\ln \beta\,, \label{eq:feq}\\     &&\frac{1}{4}(\mu')^2 = k^2 (\partial_r\ln \beta )^2\,, \qquad  
k^{-2} = l_1+\frac{1}{2}l_2 s + \frac{1}{2}l_3 q + \frac{3}{8}q - s - 3 \,. \label{eq:k2}
\end{eqnarray}

If the quantities $q$ and $s$ are satisfied to the condition, 
\begin{equation}
\frac{3}{8}q - s - 3 = 0\,, \label{eq:qs}
\end{equation}
then the equations (\ref{eq:feq}), (\ref{eq:k2}) have solutions,
\begin{equation}
\mu (r) = \frac{r_0}{r}\,, \qquad \beta (r) = \beta_0\,e^{\pm\frac{kr_0}{2r}}\,, 
\qquad k=\sqrt{k^2} >0\,, \label{eq:sol1}
\end{equation}
which leads to the metrics,
\begin{eqnarray}
&&ds^2=e^{\mp\frac{kr_0}{r}}ds^2_{YR}\,,\label{eq:metr1} \\ 
&&ds^2_{YR} = e^{-\frac{r_0}{r}}dt^2-e^{\frac{r_0}{r}}(dr^2+r^2(d\theta^2+\sin^2{\theta}\,d\phi^2))\,. \label{eq:YR}
\end{eqnarray}

With the help of the conformal transformation 
\begin{equation}
\label{eq:conf}
\check g_{\alpha\beta} = e^{\mp\frac{kr_0}{r}}g_{\alpha\beta}\,, 
\end{equation}
the metrics (\ref{eq:metr1}) can be transformed to the metrics (\ref{eq:YR}), the Cartan--Weyl space being transformed to the Riemann--Cartan space. 

If one puts $r_0 = r_g = 2Gm/c^2$, the metrics (\ref{eq:YR}) is known as the Yilmaz--Rosen (YR) metrics \cite{Yilmaz, Rosen, Itin}. In this case this metrics in the post-Newtonian approximation  at large distances gives the same results as the Schwarzschild metrics. The metrics (\ref{eq:YR}) belongs to the Majumbar--Papapetrou class of metrics \cite{Madj, Papap}. 

The metrics (\ref{eq:metr1}) will be named as the  generalized  Yilmaz--Rosen metrics. In the simplest case the constant $k$ can be chosen as $k = 1/\sqrt{l_1}$, where $l_1$ is the coupling constant in the Lagrangian density (\ref{eq:LagrPL2}). 

\section{Possible influence of dark matter \\ on the interplanetary spacecraft motion}

Let's consider a radial motion of a test body under the influence of the metrics (\ref{eq:metr1}). The $t$-component of the geodesic equation has the first integral, 
\begin{equation}
\label{eq:fint}
e^{-(1\pm k)\frac{r_g}{r}} \frac{cdt}{ds} = E_0 = const\,. 
\end{equation}
Let us divide (\ref{eq:metr1}) on $ds^2$ and put $d\theta = 0, \quad d\phi =0$. Then after some transformations we shall obtain for radial movement the following 
functional dependence between the velocity $v$ of a test body and the radial coordinate $r$, 
\begin{equation}
\label{eq:vr}
\frac{v^2}{c^2} = e^{-\frac{2r_g}{r}} \left (1-\frac{1}{E^2_0} e^{-(1\pm k)\frac{r_g}{r}} \right ) \,. 
\end{equation}
This equation yields the identity,
\begin{equation}
\label{eq:vinf}
\frac{v^2_{\inf}}{c^2} = 1 - \frac{1}{E^2_0} \,, 
\end{equation} 
where $v_{\inf}$ is an asymptotical value of the test body velocity at infinity. 

Let's apply the equalities (\ref{eq:vr}) and (\ref{eq:vinf}) to motion  of interplanetary spacecrafts starting from the Earth of radii $R$ with the velocity $v_0$(ignoring its rotation). If we use the Newton approximation in this case, then we obtain the approximate equality for the Earth,
\begin{equation}
\label{eq:vinfv0}
\frac{v^2_{\inf}} {c^2} \approx \left (1 + 3\frac{r_g}{R}\right ) 
\frac {v^2_{0}}{c^2} - (1 \pm k) \frac{r_g}{R} \,.  
\end{equation}
One can derive from \ref{eq:vinfv0}, 
\begin{equation}
\label{eq:dvinf}
\frac{v^2_{\inf} - v^2_{\inf/0}}{c^2} \approx \mp \left(\frac{r_g}{R}\right)_{Earth} \,, 
\end{equation}
where $v_{\inf/0}$ is the value of the test body velocity at infinity calculated under the condition $k = 0$. 

The data on Galileo, Cassini and other Earth flyby of the interplanetary spacecrafts shows the increases $\Delta v_{\inf}$ in the asymptotic line-of-sight velocity $v_{inf}$, of the order of $1 \div 10\; \frac{mm}{s}$ \cite{Iorio}. Therefore the value of (\ref{eq:dvinf}) is not zero. From this fact one can made two conclusions. First, we need to choose the second sign in the solutions (\ref{eq:sol1}), (\ref{eq:YR}). Second, the formula (\ref{eq:dvinf}) allows to estimate the values of $k$ and $l_1$ in the following manner.

Let us suppose that the starting velocity is equal to the second cosmic velocity (escape velocity), 
\begin{equation}
v_o \approx v_{scv}\,, \qquad \left (\frac{v_{scv}}{c^2}\right )^2 = \left(\frac{r_g}{R}\right)_{Earth}\,. \label{eq:scv}
\end{equation}
Then one can find from (\ref{eq:vinfv0}),
\begin{equation}
\label{eq:3rg}
\frac{v^2_{\inf/0}}{c^2} \approx 3\left(\frac{r_g}{R}\right)^2_{Earth} \,, 
\end{equation}
and therefore the following estimation is valid,
\begin{equation}
\label{eq:kest}
\Delta v_{\inf} = v_{\inf} - v_{\inf/0} \approx \frac{kc}{2\sqrt{3}} \approx 10^8 k \frac{m}{s}\,, \qquad  k \approx 10^{-11} \div 10^{-10}\,.
\end{equation} 

\section{Conclusions}
As a consequence of the Poincar{'}e--Weyl gauge theory of gravitation, the Dirac scalar field, which has an equally fundamental status as the metrics, should exist in Nature, and spacetime has a geometrical structure of the Cartan--Weyl space.  Such gravitational theory we have named a Weyl--Dirac theory of gravitation. In this theory we derive a static spherically symmetric solution of the field equations in vacuum for a central mass. With this solution we consider a radial motion of an interplanetary spacecraft starting from the Earth. Using the Newton approximation we obtain that the asymptotic line-of-sight velocity $v_{\inf}$ in this case depends from the parameter $k$ of the solution. Using the observable data, one can obtain the value of this parameter. 

The results were obtained within the framework of performance
of the Task No 3.1968.2014/K.

\renewcommand{\refname}{References}

\end{document}